# Distinct surface and bulk charge density waves in ultrathin 1T-TaS$_2$

Rui He[1*], Junichi Okamoto[2], Zhipeng Ye[1], Gaihua Ye[1], Heidi Anderson[1], Xia Dai[3], Xianxin Wu[3], Jiangping Hu[4], Yu Liu[5], Wenjian Lu[5], Yuping Sun[5,6,7], Abhay N. Pasupathy[8], and Adam W. Tsen[9*]

[1]Department of Physics, University of Northern Iowa, Cedar Falls, Iowa 50614, USA
[2]Department of Physics, University of Hamburg, D-20355 Hamburg, Germany
[3]Institute of Physics, Chinese Academy of Sciences, Beijing 100190, People's Republic of China
[4]Department of Physics and Astronomy, Purdue University, West Lafayette, Indiana 47907
[5]Key Laboratory of Materials Physics, Institute of Solid State Physics, Chinese Academy of Sciences, Hefei 230031, People's Republic of China
[6]High Magnetic Field Laboratory, Chinese Academy of Sciences, Hefei 230031, People's Republic of China
[7]Collaborative Innovation Centre of Advanced Microstructures, Nanjing University, Nanjing 210093, People's Republic of China
[8]Department of Physics, Columbia University, New York, New York 10027, USA
[9]Institute for Quantum Computing and Department of Chemistry, University of Waterloo, Waterloo, Ontario N2L 3G1, Canada

[*]Correspondence to: awtsen@uwaterloo.ca, rui.he@uni.edu

**Abstract:**

We employ low-frequency Raman spectroscopy to study the nearly commensurate (NC) to commensurate (C) charge density wave (CDW) transition in 1T-TaS$_2$ ultrathin flakes protected from oxidation. We identify new modes originating from C phase CDW phonons that are distinct from those seen in bulk 1T-TaS$_2$. We attribute these to CDW modes from the surface layers. By monitoring individual modes with temperature, we find that surfaces undergo a separate, low-hysteresis NC-C phase transition that is decoupled from the transition in the bulk layers. This indicates the activation of a secondary phase nucleation process in the limit of weak interlayer interaction, which can be understood from energy considerations.

Many layered, transition metal dichalcogenides (TMDs) form charge density waves (CDWs), whereby the conduction electrons and atoms displace periodically [1]. The precise mechanisms responsible for this ordering remain unresolved, although both electronic and structural instabilities are understood to play a role [2]. In 1T-TaS$_2$ alone, several CDW phases exhibiting increasingly insulating behavior appear with decreasing temperature, separated by first-order transitions. Upon cooling from the normal state, a bulk crystal first shows a CDW at 545K that is incommensurate with the lattice. In the nearly commensurate (NC) phase at 353K, the CDW forms a domain structure with locally commensurate regions. Below 183K, the domain walls disappear and the CDW becomes fully commensurate (C) [3]. Despite the highly two-dimensional (2D) structure of 1T-TaS$_2$, the CDWs in adjacent layers interact, giving them three-dimensional (3D) character [4–6].

The effects of dimensionality and interlayer coupling on the different CDW phases are areas of great current interest [7–13], which we can study directly by reducing sample thickness. Recently, several of us have shown that in ultrathin flakes produced by mechanical exfoliation [14], the C phase becomes more conducting, while NC-C transition becomes more metastable. It was suggested that reduced dimensionality enhances the pinning of conducting NC domain walls, and thus increases the activation barrier between the NC and C states. Previous measurements, performed using transport and transmission electron microscopy, however, do not distinguish between the CDWs within the bulk and on the surface, which may experience different energies. Here, we use temperature-dependent Raman spectroscopy to probe the low-frequency phonons of few-layer 1T-TaS$_2$ in the CDW state. The technique is sensitive to both bulk and surface modes, which become distinct in thin samples. We find that while the NC-C transition for bulk layers show increasing metastability for decreasing thickness, the surface transition always exhibits low activation energy. Since bulk and surface CDW transitions are identical in thick crystals [3,5,15,16], this suggests that the strength of interlayer interactions is reduced in the ultrathin limit, allowing the surface layers to decouple and undergo a separate nucleation process.

The main panel of Fig. 1(a) shows an optical image of a representative sample. In order to avoid the effects of surface oxidation [14], 1T-TaS$_2$ was exfoliated onto a silicon wafer within a nitrogen-filled glovebox, and then capped with thin hexagonal boron nitride (hBN) before transfer out to the ambient environment. A side-view schematic is shown in the inset above. This procedure is crucial, as previous works on unprotected samples prepared in air report the absence of any charge order in thin layers [17–19]. As the hBN conforms to the underlying material or substrate, an atomic force microscope was used to measure the thickness of the buried 1T-TaS$_2$ post-transfer (see Supplementary Material). In Fig. 1(b), we show Raman spectra between 40 and 120 cm$^{-1}$ of an 8.1 nm thick flake for a series of temperatures upon cooling, taken with 532 nm laser light focused to a spot size of ~ 2 μm. In the NC phase at high temperature (> 150 K), several broad peaks are observed, with two intense peaks centered at ~ 70 and ~ 76 cm$^{-1}$. In the low-temperature C phase (< 150 K), many additional peaks appear that are well-resolved down to less than one wavenumber. These features are in close agreement with previous studies on thick crystals [20,21], indicating that the 8.1 nm flake possesses mostly bulk-like characteristics.

In order to determine the precise temperature at which the NC-C phase transition takes place, we plotted the Raman frequencies for each discernable peak as a function of temperature for cooling (Fig. 1(c)) and warming (Fig. 1(d)). Clear changes in both the number of modes as well as their frequencies appear at the transition temperature $T_c$ (marked by the dashed line), which is different for temperature sweeps down ($T_{c,cool}$ = 140 K) and up ($T_{c,warm}$ = 210 K), as expected for a strong first-order phase transition. The mode at ~ 110 cm$^{-1}$ (colored red) does not show strong changes at $T_c$ and is attributed to a surface mode, the details of which shall be discussed in the sections to follow. We obtain a hysteresis temperature of $\Delta T = T_{c,warm} - T_{c,cool} = $ 70 K and an average transition temperature of $T_{c,avg} = (T_{c,warm} + T_{c,cool})/2 = 175$ K, similar to that recently observed in exfoliated flakes of comparable thickness using transport measurements [14]. The effect of laser-induced heating on the transition temperature is discussed in the Supplementary Material.

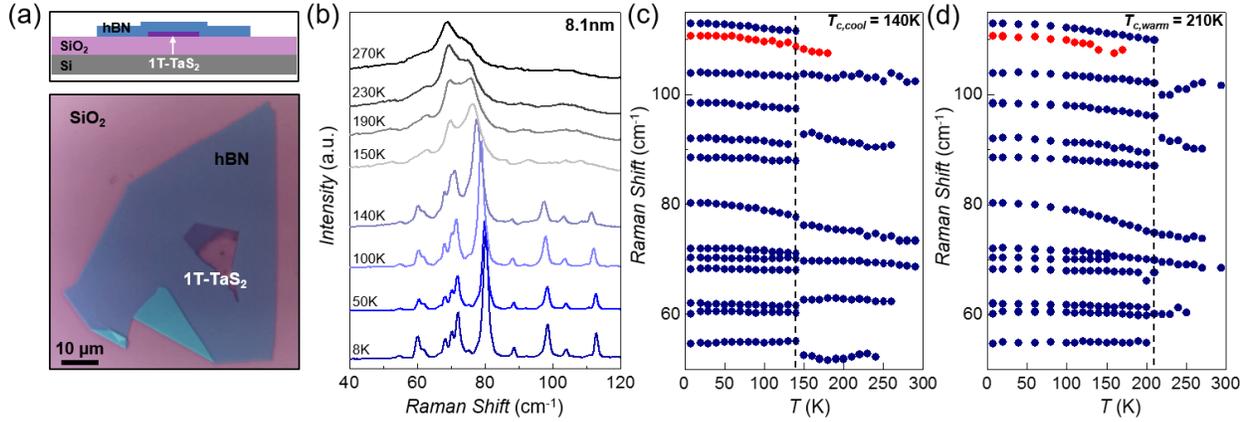

FIG. 1 (a) Optical image of a representative 1T-TaS$_2$ thin flake sample, protected by hBN. Cross-sectional schematic is shown above. (b) Low-frequency Raman spectra of an 8.1-nm-thick flake at different temperatures in the cooling process. Frequencies of discernible peaks are plotted as a function of (c) decreasing and (d) increasing temperature. NC-C CDW transition temperature is marked by a dashed line and shows hysteresis between cooling and warming. Mode colored red is due to surface CDW and to be discussed later.

We now turn to the thickness dependence of the Raman spectra. In Fig. 2, we show measurements for three flakes of different thicknesses (1.5, 4.3, and 8.1 nm—corresponding to 3, 7, and 14 layers, respectively), as well as that of a bulk crystal, at both (a) ~ 250 K (NC phase) and (b) ~ 10 K (C phase). The spectra taken in the NC phase are similar for all four samples, although the peaks at ~ 60 and ~ 75 cm$^{-1}$ are slightly more pronounced for the exfoliated flakes. This suggests that the structure of the NC CDWs do not fundamentally change with reduced dimensionality. In contrast, large changes are observed in the C phase. The features generally broaden with decreasing thickness and closely spaced peaks seen in the crystal at low wavenumbers (below 80 cm$^{-1}$) can no longer be resolved in the thinnest flake, which could be a consequence of increased disorder [22].

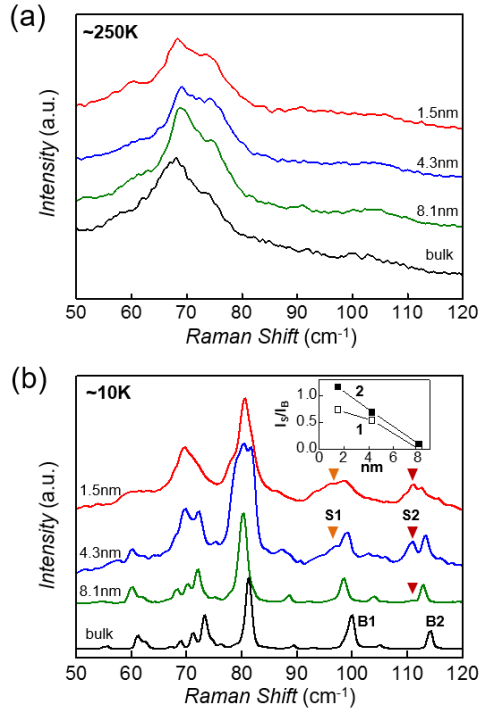

FIG. 2. Raman spectra for different thickness flakes along with bulk crystal taken at (a) 250 K and (b) 10 K. Orange and red arrows mark surface modes. Intensity ratio of corresponding surface and bulk modes ($I_{S1}/I_{B1}$ and $I_{S2}/I_{B2}$) as a function of thickness is plotted in inset of (b).

At the same time, new peaks appear in thinner samples, the most discernible at 97 and 111 cm$^{-1}$ are marked using orange and red triangles, respectively. In a recent study of 1T-TaS$_2$, several effects were predicted to change the C phase Raman characteristics of thin samples [11]. First, depending on whether the number of layers is even or odd, calculations show large differences in both the number of modes and their positions. Yet, the observed spectra are qualitatively similar for samples with different parity layers. Second, sample strain may induce peak shifts; however, it does not produce additional Raman modes. Third, a change in c-axis CDW stacking from hexagonal to triclinic increases the number of modes, but this also leaves a gap around 100 and 110 cm$^{-1}$. Thus, the new peaks cannot be explained by these effects. Finally, as will be shown below, the reduced temperature hysteresis between cooling and warming for these modes are also inconsistent with the scenario of disorder-activated peaks, since disorder tends to either suppress the NC to C transition, or increase the temperature hysteresis [23,24].

Instead, the growing intensity of these modes with lower thickness suggests that they originate from surface phonons. Their close proximity to existing peaks further suggests that they share the same vibrational character as the original bulk modes, although with slightly different energy. In the inset of Fig. 2(b), we plot the ratio of intensities between the two peak couples (S1 and B1; S2 and B2) as a function of thickness. The monotonic increase of $I_S/I_B$ with decreasing flake thickness indicates that S1 (S2) is likely a surface phonon mode of the same character as mode B1 (B2) in the bulk layers.

In earlier work on bulk 1T-TaS$_2$ [21], modes B1 and B2 were identified as C phase acoustic phonon modes arising primarily from the vibration of Ta atoms. While one may expect such a mode to soften in thin samples as interlayer interactions disappear, the growth in intensity of secondary surface modes that are separately resolved from the bulk modes is unique. It indicates that the surface CDW in the C phase (S modes) becomes spectroscopically distinct from that within the bulk layers (B modes), which could result from a decoupling of the CDW on the outermost layers.

In order to better understand these results as well as the difference between the NC and C phases, we have carefully studied the temperature evolution of the S and B modes across the NC-C transition. In Fig. 3, we show Raman spectra between ~ 95 to ~ 115 cm$^{-1}$ upon (a) cooling and

(b) warming for the 4.3-nm-thick flake, in which all four peaks are most clearly resolved. We observe that the S peaks appear at a higher temperature than the B peaks during cooling, but disappear at lower temperature during warming. This indicates that the bulk and surface CDWs undergo separate NC-C transitions. Earlier studies on thick 1T-TaS$_2$ crystals using separate bulk and surface probes have reported nearly identical transition temperatures for the different measurements [3,5,15,16]. Hence, the separation of the surface and bulk NC-C transitions is unique to thin samples. We note that changes in the lower energy modes across the phase transition are less well-resolved in thin samples (see Supplementary Material).

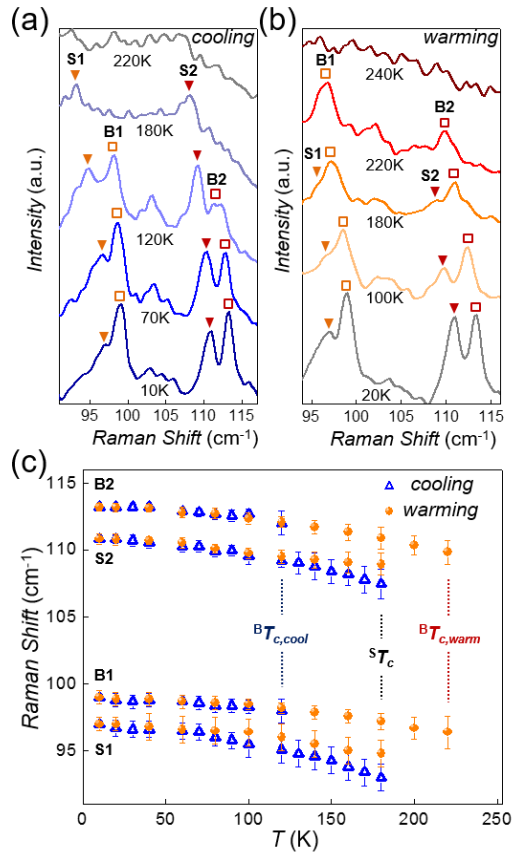

FIG. 3. Temperature evolution of peaks in the 95-115 cm$^{-1}$ region for 4.3-nm-thick flake during (a) cooling and (b) warming. Surface modes S1 and S2 appear (disappear) before bulk modes B1 and B2, respectively, when cooling (warming). (c) Measured Raman frequencies vs. temperature for cooling and warming. The error bars are obtained from fitting the peaks at different temperatures to a Lorentzian lineshape. Surface modes show same transition temperature in both directions while bulk modes show large temperature hysteresis.

When each peak can be clearly distinguished from the background, we have measured their position as a function of temperature, and the results are shown in Fig. 3(c) for both cooling and warming. Each bulk and surface mode pair show similar softening with increasing temperature, further substantiating that they originate from the same phonon vibration (see Supplementary Material). The bulk modes B1 and B2 appear at $^{B}T_{c,cool}$ = 120 K when decreasing temperature and disappears above $^{B}T_{c,warm}$ = 220 K when increasing temperature, which gives a temperature hysteresis of $^{B}\Delta T$ ~ 100 K and $^{B}T_{c,avg}$ ~ 170K. The surface modes S1 and S2, however, appears at (and disappears above) $^{S}T_{c,}$ = 180 K, independent of the direction of temperature change. This hysteresis-free value is close to $^{B}T_{c,avg}$.

We have also measured the bulk and surface transition temperatures for the other flakes similarly, and the combined data are shown in Fig. 4(a) and (b) as a function of sample thickness. With lower thickness, $^{B}T_{c,cool}$ (blue circles) decreases and $^{B}T_{c,warm}$ (orange circles) increases by similar amounts. Thus, $^{B}\Delta T$ grows with decreasing thickness, while $^{B}T_{c,avg}$ stays nearly constant. The blue and orange dashed lines mark the respective cooling and warming transition temperatures measured for the bulk crystal. These results are consistent with previously measured transport properties [14], which are likely determined by the bulk layers for samples greater than a few layers thick. In contrast, cooling and warming transition temperatures for the surface modes from all the three samples are nearly the same within the experimental error and remain close to $^{B}T_{c,avg}$. This indicates that in ultrathin samples, only the bulk layers experience increased metastability, while the decoupled surface layers undergo the NC-C transition at $^{S}T_c$ ~ $^{B}T_{c,avg}$.

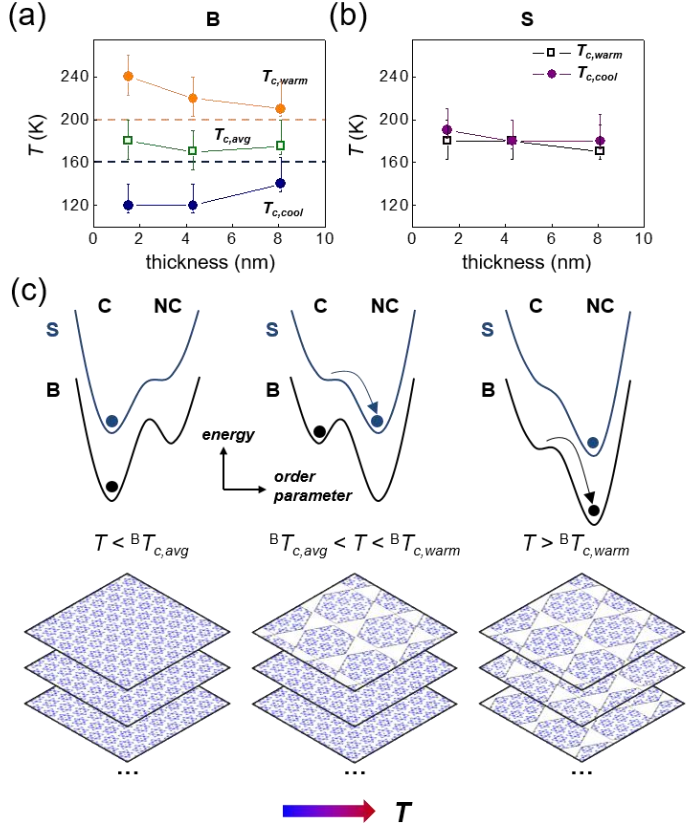

FIG. 4. Transition temperatures as a function of flake thickness for (a) bulk and (b) surface modes. Thinner samples show larger temperature hysteresis for the bulk modes. Transition for surface modes is nearly constant with thickness and occurs close to average bulk transition temperature. The error bars are determined by our temperature steps (lower bound) as well as the temperature increase from laser heating (upper bound) (see Supplementary Material). (c) Free energy (above) and real space schematics (below) showing separate bulk and surface CDW phase transitions during warming. Above $^{B}T_{c,avg}$, surface layers are decoupled and make separate transition into NC phase.

These effects can be summarized with reference to the diagrams in Fig. 4(c) describing the warming transition as an example. We show real space schematics of the temperature evolution of few-layer 1T-TaS$_2$ on the lower set of panels and corresponding free energy diagrams above. The metastability of a phase transition reflects the activation barrier separating states of free energy minima. Upon warming from low temperature, the entire sample starts in the C phase ground state for $T < {}^{B}T_{c,avg}$. When raising temperature to $^{B}T_{c,avg} < T < {}^{B}T_{c,warm}$, the NC phase becomes the thermodynamic ground state. Here, the bulk interior layers do not have the necessary energy to overcome the activation barrier and remains in the C state. Since the surface layers are decoupled, however, they undergo a separate phase transition to the NC phase as domain walls nucleate and grow in those layers. At $T > {}^{B}T_{c,warm}$, the activation barrier has become small relative to the thermal energy, and the bulk layers finally also transition into the NC phase. The order of the transitions is reversed during cooling.

The larger activation barrier in thinner flakes has been attributed to enhanced pinning of nucleated domain walls by impurity centers [14]. The low hysteresis on the surface, however, then suggests that the phase nucleation mechanism for the NC-C transition is fundamentally different when the CDWs are decoupled between layers. We have performed an energy analysis of different nucleation processes for NC formation within the C phase [25-27], which we describe in brief below. Details can be found in the Supplementary Material. Our results can be understood as a crossover from 3D to 2D phase nucleation as interlayer interactions are reduced.

Fig. 5(a) shows schematic structures of two likely NC critical-size nuclei. It is possible for a disk of the NC phase to form within an individual layer only (2D nucleation), or in every layer, stacking together coherently (3D nucleation). The total energy of a 3D nucleus is the energy cost of a single NC disk of size $R$, $E_{disk}(R)$, multiplied by number of layers $N$: $E_{3D} = NE_{disk}(R)$. For 2D nuclei, the absence of disks in adjacent layers costs additional energy proportional to its area due to the loss of favorable NC interlayer interactions, which we model as: $E_{2D} = E_{disk}(R) + J_c R \xi / d^2$, where $J_c$ represents the interlayer CDW coupling strength, $\xi$ is the characteristic width of domain walls, and $d$ is the interlayer separation distance. While 3D nucleation is favored when interlayer CDW interactions are strong, 2D nuclei require less energy as $J_c$ vanishes.

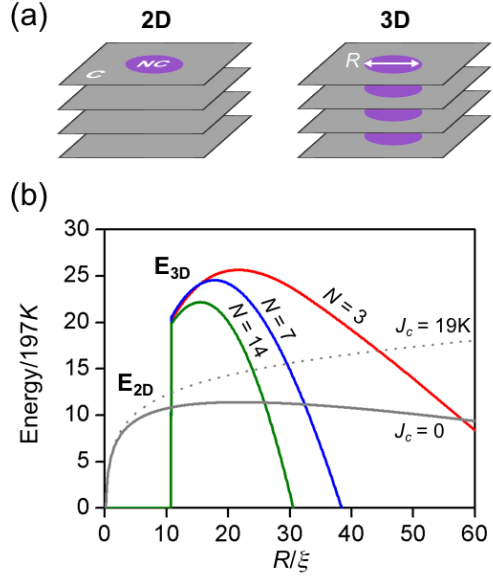

FIG. 5. (a) 2D and 3D NC phase nucleation model corresponding to surface and bulk transitions, respectively. $R$ is the size of the NC phase critical nucleus. (b) Energy as a function of $R$ normalized to characteristic width of domain walls $\xi$. Energy analysis shows larger activation barrier for thinner samples in 3D nucleation process. 2D nucleation becomes more favorable with vanishing interlayer CDW coupling.

Impurities and defects have been observed in even nominally pure 1T-TaS$_2$ samples [12,13], and so one must further take into account the effects of CDW pinning [28]. Pinning centers hinder the growth of 2D NC nuclei and can be described by an additional energy term: $E_{disk} \rightarrow E_{disk} + E_{pin}$. We expect $E_{pin} \sim N^{-2/3}$ in the regime of weak, collective pinning for moderately anisotropic cases [14], while for extremely thin samples, it approaches the individual pinning limit $E_{pin} = n_{imp}\xi$. In Fig. 5(b), we have plotted $E_{3D}$ vs. $R/\xi$ for $N = 3$, 7, and 14, corresponding to the number of layers in our different thickness samples (1.5, 4.3, and 8.1 nm, respectively) and for appropriately chosen material parameters. In all three curves, $E_{3D}$ shows an activation energy maximum $E_{3D,max}$ for a critical nucleus size and decreases to negative values as $R$ is increased further. $E_{3D,max}$ increases for decreasing $N$, consistent with the large hysteresis observed for the bulk modes. Also in Fig. 5(b), we show plots of $E_{2D}$ vs. $R/\xi$ for two different values of interlayer coupling. For large $J_c$, $E_{2D}$ grows monotonically with $R$, while for $J_c = 0$, $E_{2D}$ decreases at large $R$, with $E_{2D,max}$ less than half $E_{3D,max}$.

These results show that the formation of 2D nuclei is generally unfavorable, but becomes the dominant (less costly) nucleation process in the limit of vanishing interlayer coupling. We thus identify 3D nucleation as the highly metastable process occurring in the well-coupled, bulk layers of ultrathin 1T-TaS$_2$ and attribute the low-hysteresis transition to 2D phase nucleation realized on the decoupled surface layers. It should be noted that C phase Raman modes have been recently observed in a monolayer 1T-TaS$_2$ sample at low temperature [11], indicating that the commensurate state remains the thermodynamic ground state and is obtainable in the single layer limit. This observation is consistent with our scenario for a low hysteresis surface transition, although further temperature dependent studies on monolayer samples are needed to confirm this. Finally, it has recently been proposed that changing $c$-axis orbital ordering may be used to tune the in-plane electronic structure of 1T-TaS$_2$ in the C phase [29]. The surface decoupling we observe in ultrathin flakes may potentially allow for a similar device concept, whereby controlling interlayer coupling in a bilayer sample can switch between layer independent and interdependent conduction.

**Acknowledgments:**

We thank T. Kidd (Iowa) and G. Sciaini (Waterloo) for helpful discussions, and J. Shi (Harvard) for assistance with sample preparation. Work at the University of Northern Iowa is supported by the National Science Foundation (NSF, Grants No. DMR-1552482 (RH and GY) and DMR-1410496 (RH and ZY)) and the American Chemical Society Petroleum Research Fund (Grant No. 53401-UNI10 (HA)). R.H. also acknowledges support by the UNI Faculty Summer Fellowship. The low temperature equipment was acquired through the NSF MRI Grant (No.

DMR-1337207). Work at Columbia University is supported by the NSF via Grant. No. 1124894 (AWT) and DMR-1056527 (ANP). Work in China was supported by the National Key Research and Development Program (2016YFA0300404), the National Nature Science Foundation of China (11674326, 11404342), the Joint Funds of the National Natural Science Foundation of China and the Chinese Academy of Sciences' Large-scale Scientific Facility (U1232139).